\documentclass[twocolumn,showpacs,preprintnumbers,amsmath,amssymb]{revtex4}

\usepackage{graphicx}
\usepackage{dcolumn}
\usepackage{bm}
\usepackage{graphics}
\usepackage{latexsym}
\usepackage{graphics}
\usepackage{dcolumn}
\usepackage{epsfig}
\usepackage{amssymb} 
\usepackage{amsmath} 
\usepackage{rotating} 
\usepackage{rotate} 
\usepackage{color} 
\usepackage{graphicx}
\usepackage{amssymb}

\newcommand{\gp}{\dot{\gamma}}

\begin{document}
\title{Shear-induced structuration of confined carbon black gels:\\ 
Steady-state features of vorticity-aligned flocs}
\author{Vincent Grenard}
\affiliation{Universit\'e de Lyon, Laboratoire de Physique, \'Ecole Normale Sup\'erieure de
Lyon, CNRS UMR 5672, 46 All\'ee d'Italie, 69364 Lyon cedex 07, France}
\author{Nicolas Taberlet}
\affiliation{Universit\'e de Lyon, Laboratoire de Physique, \'Ecole Normale Sup\'erieure de
Lyon, CNRS UMR 5672, 46 All\'ee d'Italie, 69364 Lyon cedex 07, France}
\author{S\'ebastien Manneville}
\affiliation{Universit\'e de Lyon, Laboratoire de Physique, \'Ecole Normale Sup\'erieure de
Lyon, CNRS UMR 5672, 46 All\'ee d'Italie, 69364 Lyon cedex 07, France}
\email{sebastien.manneville@ens-lyon.fr}
\date{\today}

\begin{abstract}
Various dispersions of attractive particles are known to aggregate into patterns of vorticity-aligned stripes when sheared in confined geometries. We report a thorough experimental investigation of such shear-induced vorticity alignment through direct visualization of carbon black gels in both simple plane shear and rotational shear cells. Control parameters such as the gap width, the strain rate, and the gel concentration are systematically varied. It is shown that in steady states the wavelength of the striped pattern depends linearly on the gap width $h$ while being insensitive to both the gel concentration $C$ and the shear rate $\gp$. The width of the vorticity-aligned flocs coincides with the gap width and is also independent of $C$ and $\gp$, which hints to a simple picture in terms of compressible cylindrical flocs. Finally, we show that there exists a critical shear rate $\gp_c$ above which structuration does not occur and that $\gp_c$ scales as $h^{-\alpha}$ with $\alpha=1.4\pm 0.1$ independently of $C$. This extensive data set should open the way to quantitative modelling of the vorticity alignment phenomenon in attractive colloidal systems.
\end{abstract}
\maketitle

\section{Introduction \label{s.intro}}

Colloidal gels made of attractive particles at low volume fraction raise both fundamental and industrial interest \cite{Russel:1989,Larson:1999}. For instance, among the huge variety of colloidal systems, those made of fumed carbon black particles suspended into an organic solvent or matrix are involved in many applications ranging from paints, coatings and printing to rubbers and tires \cite{Donnet:1993}. Carbon black gels have also raised academic interest in the context of jamming \cite{Trappe:2000,Trappe:2001}, shear-thickening \cite{Osuji:2008b,Negi:2009a}, and dynamical heterogeneity as well as aging in attractive systems \cite{Trappe:2007,Negi:2009b}. Besides their macroscopic rheological and aging properties, recent research effort has focused on the effect of confinement on such systems in search of new prospects in miniaturization and microfluidic applications. In particular, it was shown that, when sheared within a confined geometry, various attractive particulate systems display a striking structuration into a striped pattern of log-rolling flocs aligned along the vorticity direction \cite{Vermant:2005}. Vorticity alignment was both inferred from light scattering experiments in colloid--polymer mixtures \cite{DeGroot:1994} and Laponite clay suspensions \cite{Pignon:1996} and directly demonstrated by structure visualization in flocculated magnetic suspensions \cite{Navarrete:1992,Navarrete:1996}, non-Brownian carbon nanotubes \cite{Lin-Gibson:2004}, attractive emulsions \cite{Montesi:2004}, and carbon black and alumina dispersions \cite{Negi:2009b,Osuji:2008a}, hinting to a somewhat general phenomenon.

A mechanism for vorticity alignment was proposed based on an elastic instability of soft viscoelastic domains embedded into a less viscoelastic fluid: due to the streamline curvature inside the domains, an internal ``hoop stress''  develops that compresses the domains in the radial direction, leading to an elongation along the vorticity direction \cite{Lin-Gibson:2004,Montesi:2004}. This scenario is supported by the observation of small but noticeable negative first normal stress differences \cite{Montesi:2004,Lin-Gibson:2004,Osuji:2008a}. Although this effect, similar to a Weissenberg effect localized within viscoelastic domains, may indeed be involved in two-phase viscoelastic fluids  \cite{Hobbie:1999,Hobbie:2004}, it is less clear how compressible aggregates of attractive particles may play the role of viscoelastic domains. Moreover, such a qualitative scenario does not predict the features of the striped pattern such as its wavelength and the width of the stripes. From the experimental point of view, the current literature suffers from a lack of detailed, quantitative characterization of the vorticity alignment phenomenon and only rather qualitative and fragmented data is available on the pattern static and dynamic features. In the particular case of carbon black gels, the vorticity-aligned structures were investigated only in rotational shear devices, such as the plate-plate and the cone-and-plate geometries, where either the shear rate or the gap width is spatially inhomogeneous \cite{Osuji:2008a,Negi:2009a}. Therefore, the aim of the present paper is to complement previous results on carbon black gels with a thorough analysis of vorticity-aligned patterns in various geometries, including simple plane shear, by varying all the control parameters (gel concentration, shear rate, and gap width).

The present article is devoted to the steady-state characteristics of vorticity-aligned flocs. In Sect.~\ref{s.matmeth} below, we first briefly explain the preparation protocol, recall the basic rheological characteristics of carbon black gels, and describe the translational and rotational setups used to shear the samples within gaps of width $h=40$--400~$\mu$m. The experimental results are presented in Sect.~\ref{s.results}, where we first qualitatively compare the patterns observed in three different geometries before turning to a quantitative image analysis of the influence of the various control parameters. Finally our results, which should feed future modelling of vorticity alignment in attractive systems, are discussed in Sect.~\ref{s.discuss} in terms of compressible cylindrical flocs.

\section{Materials and methods \label{s.matmeth}}

\subsection{Carbon black gels \label{s.cb}}

\subsubsection{Sample preparation.~~\label{s.prepa}}

\begin{figure}[htb]
\begin{center}
\scalebox{1.1}{\includegraphics{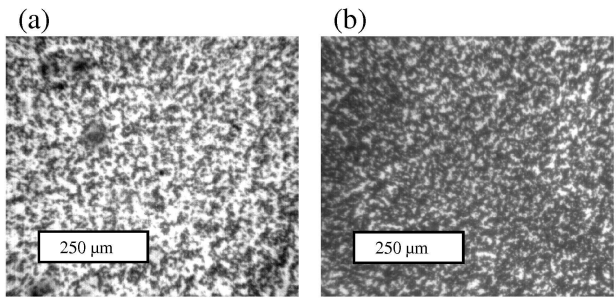}}
\end{center}
\caption{\small{Optical microscopy images of carbon black gels at (a)~1~\% w/w and (b)~2~\% w/w.} }
\label{f.structure}
\end{figure}

Our gels consist of dispersions of carbon black (CB) particles (Cabot Vulcan XC72R) in a light mineral oil (Sigma, density 0.838, viscosity 20~mPa.s) at concentrations ranging from 0.25 to 3~\% w/w as described by Trappe {\it et al.} \cite{Trappe:2007} CB particles (density 1.8, typical diameter 200~nm, fractal dimension 2.2) are themselves small aggregates of permanently fused ``primary'' particles of diameter 20--40~nm. Our samples are prepared by dispersing the CB powder into the oil through a vigorous manual agitation followed by a 1 hour sonication, which breaks any aggregates between CB particles. CB particles interact through a short-range attractive potential, whose depth is estimated at about 30~$kT$ for a 2~\% w/w suspension \cite{Trappe:2007}. This attraction results in the formation of a weak colloidal gel that is easily disturbed by shear. Figure~\ref{f.structure} shows the structure of $\sim$30 $\mu$m-thick samples as seen under a standard optical microscope. As previously shown by Trappe {\it et al.} \cite{Trappe:2000,Trappe:2001}, the gel is formed by a network of interconnected CB particles, which gets denser as the concentration is increased.

\subsubsection{Rheological properties.~~\label{s.rheol}}

The rheological measurements presented in this paragraph are performed at room temperature ($\sim$25$^\circ$C) with a stress-controlled rheometer (Anton Paar, MCR 301) in a Plexiglas Couette cell of gap width 1~mm. In order to ensure a reproducible initial gel state, the sample is presheared at a high shear rate (+1000~s$^{-1}$ then -1000~s$^{-1}$ for 20~s each) prior to any measurement. The gel is then left at rest for 100~s, which is sufficient to let the gel structure reform and reach a steady state with negligible aging \cite{Negi:2009b,Gibaud:2010}. As discussed in previous studies \cite{Osuji:2008a,Osuji:2008b}, preshear at high shear rates induces the breakup of locally dense clusters of CB particles into less dense clusters, leading to an increase of the effective particle volume fraction and therefore to enhanced viscous dissipation i.e. shear-thickening. From this shear-thickened gel state, vorticity-aligned structures are readily formed under steady shear at low shear rates \cite{Osuji:2008a,Negi:2009a}.

\begin{figure}[htb]
\begin{center}
\scalebox{1.0}{\includegraphics{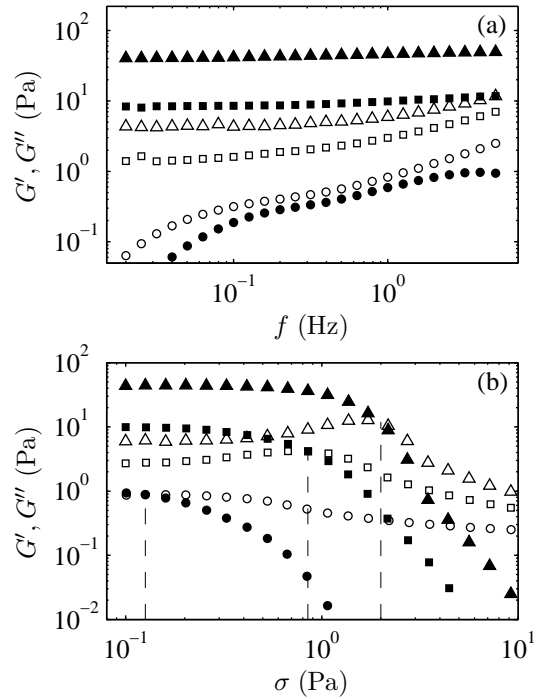}}
\end{center}
\caption{\small{Oscillatory shear rheology. Storage modulus, $G'$, (solid symbols) and loss modulus, $G''$, (open symbols) for CB gels at various concentrations: 1~\% ($\circ $), 2~\% ($\square $), and 3~\% w/w ($\triangle $). (a) $G'$ and $G''$ as a function of the frequency $f$ for a given stress amplitude of 0.2~Pa. (b) $G'$ and $G''$ as a function of the shear stress amplitude $\sigma$ for $f=1$~Hz. Dashed lines indicate the points where $G'=G''$, corresponding to the crossover from solidlike to fluidlike behaviour as $\sigma$ is increased.}}
\label{f.osc_rheol}
\end{figure}

As shown in Fig.~\ref{f.osc_rheol}(a), the frequency-dependent viscoelastic moduli of 2~\% and 3~\% w/w CB gels are typical of soft solids: the storage modulus $G'$ is almost constant and a few times larger than the loss modulus $G''$, which weakly increases with the frequency. Moreover, the stress sweeps of Fig.~\ref{f.osc_rheol}(b) allow one to roughly estimate the yield stresses of 2~\% and 3~\% w/w samples to 0.85~Pa and 2~Pa respectively. In the case of the very weak 1~\% w/w sample, the crossover from solidlike to fluidlike can be distinguished around 0.2~Pa. Correspondingly, the frequency sweep of Fig.~\ref{f.osc_rheol}(a) performed at a stress amplitude of 0.2~Pa shows that $G''\gtrsim G'$ for the 1~\% w/w gel over the whole range of frequencies. Note that these estimates should be taken with care due to possible wall slip and time-dependent effects \cite{Gibaud:2010}. Still these rheological data are compatible with those found in the literature for similar concentrations \cite{Trappe:2000,Osuji:2008b,Gibaud:2010}.

\subsection{Experimental shear cells \label{s.setups}}

\subsubsection{Simple shear geometry.~~\label{s.simple}}

\begin{figure}[htb]
\begin{center}
\scalebox{1.0}{\includegraphics{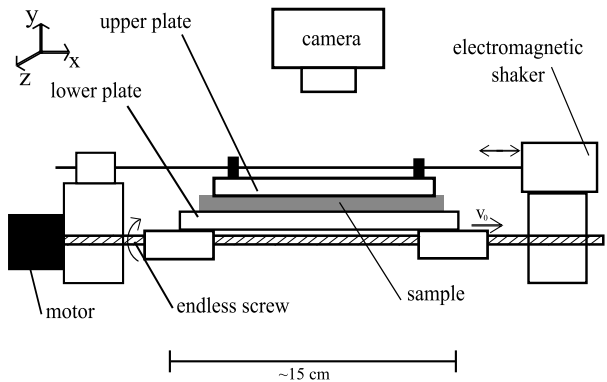}}
\end{center}
\caption{\small{Sketch of the simple shear cell as seen from the side.}}
\label{f.simple_shear_setup}
\end{figure}

In a first series of experiments, we shall use a home-made simple shear cell that consists of two parallel glass plates. This setup is sketched in Fig.~\ref{f.simple_shear_setup}. The lower glass plate can be slowly translated using an endless screw. We use a stepper motor to control the rotation speed of this endless screw and therefore the applied shear rate $\gp$. The gap width, $h$, is fixed using two thin aluminum spacers. The spacers (1~cm wide and 25~cm long) are glued over the whole length of the lower plate. A force is applied downwards on the upper plate through a metallic rod that is slighlty bent in order to force the contact between the spacers and the upper plate. By directly following the actual motion of the bottom plate using a video camera, we checked that this force is small enough so that the bottom plate moves at a constant speed $v_0$ with negligible stick-slip between the spacers and the upper plate over velocities down to 10~$\mu$m.s$^{-1}$.

As recalled above, the CB gel has to be presheared in order to get reproducible results and to reach a shear-thickened state. To this aim, the metallic rod that keeps the upper plate against the spacers is attached to an electromagnetic shaker (LDS, V201) that imposes a sinusoidal displacement at a frequency of 10~Hz and with an amplitude of 3~mm. This corresponds to an oscillatory shear rate of amplitude $\sim 1100$~s$^{-1}$. Before starting an actual experiment at low shear rate, such a large amplitude oscillatory shear is imposed for 50~s, after which the sample is homogeneously black over the whole surface of the glass plates. This indicates that any large scale heterogeneity in the gel structure has been broken up by shear. The sample is then left at rest for 10~s before the stepper motor controlling the motion of the lower plate is turned on. We checked that changing the preshear and rest durations does not affect our results. 

The gap width $h=173\pm 10~\mu$m was measured using high-frequency ultrasonic echography and further checked by measuring the apparent surface covered by small oil droplets of known volume and squeezed between the two glass plates. $h$ was found to be homogeneous over the whole surface of interest, a rectangle of typical length 150~mm and width 45~mm. This home-made parallel-plate setup, hereafter referred to as {\it translational} shear cell, allows us to shear a very large surface of sample ($\sim$70~cm$^2$) with a {\it homogeneous} strain field over distances as large as 100~mm, i.e. a total strain that can reach $\sim \gamma=60,000$~\%, with shear rates ranging from 0.1 to 3~s$^{-1}$.

In order to follow the evolution of the gel structure under shear, a CCD camera (Mikrotron MC1310) is fixed above the shear cell. This camera is used with a standard objective (Fujinon, focal length 16~mm) for wide-field imaging or together with a microscope objective (Leica, N Plan 10$\times$) to get an enlarged view of the samples (see, e.g., Fig.~\ref{f.simple_images}). The whole setup is lit by a LED backlight source (PHLOX, LEDW-BL-200$\times$200) that provides a homogeneous white lighting over 200~mm by 200~mm.


\subsubsection{Rheo-optical setup.~~\label{s.rheooptic}}

While the great advantage of our parallel plate setup is to provide a confined plane shear flow over more than 10~cm, its obvious drawbacks are that (i) it does not allow us to easily vary the gap width $h$ and (ii) it does not give access to the shear stress $\sigma$ exerted onto the sample. Indeed tiny force measurements on a large plate dragged over large distances and with the constraint of keeping the gap width constant are very difficult if not impossible. Therefore, in order to perform stress measurements, we turn to a rheo-optical setup similar to that used previously by Osuji {\it et al.} \cite{Osuji:2008a,Osuji:2008b,Negi:2009a}, although we shall not deal with the rheological measurements in the present paper.

\begin{figure}[htb]
\begin{center}
\scalebox{1.0}{\includegraphics{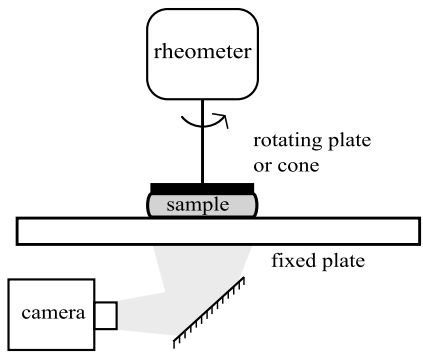}}
\end{center}
\caption{\small{Sketch of the rotational shear cell as seen from the side.}}
\label{f.rotational_setup}
\end{figure}

This rheo-optical setup, hereafter referred to as {\it rotational} shear cell, is sketched in Fig.~\ref{f.rotational_setup}. A rheometer (Anton-Paar, MCR 301) applies a shear rate $\gp$ and measures the corresponding stress response $\sigma(t)$. This stress-imposed rheometer is used in a strain-controlled mode thanks to a feedback loop that allows one to impose a constant $\gp$ within a time corresponding to a shear strain of less than 5~\% for the range of shear rates investigated here. The CB sample is visualized from below through a mirror fixed under the transparent Plexiglas lower plate using the same camera and objectives as those of the translational setup. For the larger magnification, the microscope objective is set between the lower plate and the mirror.

In order to explore the influence of the geometry, we use either a Plexiglas plate (of diameter 60~mm) or a stainless steel cone (of angle 2$^{\circ}$, diameter 50~mm, and truncation $209~\mu$m) as the upper rotating tool. Such a rotational setup does not suffer from any limitation in the accessible strain. The vertical position of the rotating tool is controlled within 1~$\mu$m. However, due to the tolerance in machining the bottom Plexiglas plate, the gap width $h$ may vary locally by $\pm 5~\mu$m so that, for the plate-plate geometry, gap widths down to $40~\mu$m only were considered. 

As recalled in the introduction, the plate-plate geometry has a homogeneous gap width but the stress and strain fields are strongly heterogeneous: the shear rate depends linearly on the radial position $r$ as $\gp(r)=r\Omega_0/h$ where $\Omega_0$ is the rotation speed of the upper plate. In the following, we shall denote by $\gp$ the shear rate at the periphery of the upper plate so that $\gp(r)=\gp r/R$ where $R$ is the radius of the upper plate. On the other hand, the cone-and-plate geometry has quasi-homogeneous stress and strain fields but the gap width varies from $h=209~\mu$m at the cone apex to $h=875$~$\mu$m at the cone periphery.

The preshear protocol is as described above in Sect.~\ref{s.rheol} for rheological measurements. The only difference is that, in the case of the plate-plate geometry, the rotation speed $\Omega_0$ (rather than the shear rate $\gp$) is kept constant whatever the gap width in order to avoid expulsion of the sample for the largest gaps. $\Omega_0$ is chosen so that $\gp=10,000$~s$^{-1}$ for $h=100~\mu$m. Although this preshear protocol differs from the oscillatory protocol used in the translational setup described above, it sets the gel in similar initial shear-thickened and reproducible states, allowing for a direct comparison between the various devices and geometries.

\section{Experimental results \label{s.results}}

In this section results obtained with the setups introduced above are described. We first give an account of qualitative observations made under simple shear and in both the plate-plate and cone-and-plate rotational geometries. We then turn to a quantitative analysis focusing on the influence of the various control parameters, namely the sample concentration $C$, the gap width $h$, and the applied shear rate $\gp$. In all cases, the data presented here are taken in ``steady-state,'' i.e. after the transient corresponding to shear-induced structuration and before any instability of the shear-induced structures \cite{Osuji:2008a}. A quantitative study of the initial transient and of long-time stability will be the subject of future work.

\subsection{Qualitative observations \label{s.qualit}}

\subsubsection{Simple shear geometry.~~\label{s.simple_images}}

\begin{figure}[htb]
\begin{center}
\scalebox{1.0}{\includegraphics{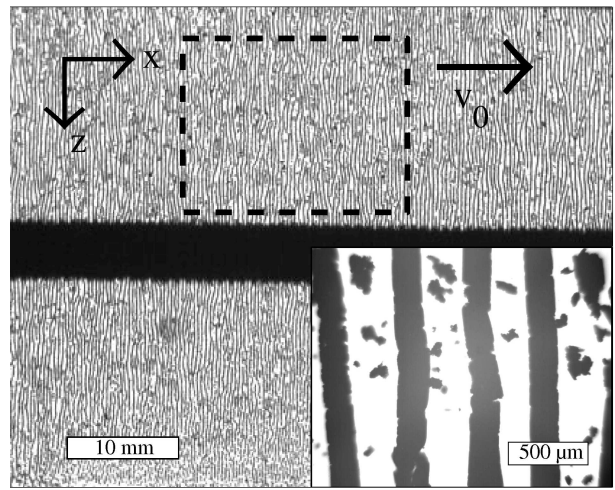}}
\end{center}
\caption{\small{Shear-induced structures under simple plane shear. The arrow denoted as $v_0$ shows the direction of shear. The dotted frame shows the typical area used for computing the wavelength. Inset: same shear-induced state at a larger magnification. These pictures were taken once the vorticity-aligned flocs are fully developed for $C=1.5$~\% w/w, $h=173~\mu$m, and $\gp=1.34$~s$^{-1}$.}}
\label{f.simple_images}
\end{figure}

Figure~\ref{f.simple_images} shows the highly-ordered shear-induced state of the gel in the translational shear cell at two different magnifications. The dark line in the middle of the wide-field image corresponds to the shadow of the endless screw. As already reported in a rotational setup by Osuji {\it et al.} \cite{Osuji:2008a,Negi:2009a}, the (opaque) CB particles arrange into long, dense flocs aligned perpendicularly to the shearing direction. The oil is transparent and fills the gaps between vorticity-aligned flocs in the main picture. With the present parallel plate setup, straight flocs with a diameter of about 100~$\mu$m are observed that can be as long as 10~mm, which corresponds to an aspect ratio of 100. The inset of Fig.~\ref{f.simple_images} also shows that some large free clusters of CB particles remain in the interstitial space between the flocs. Moreover, movies\dag~of the sample under shear at the largest magnification clearly show that the flocs rotate as solid-bodies in a log-rolling motion. In such movies, it can be seen that in this shear-induced ``steady-state,'' flocs may break and recombine with other neighboring flocs and that the oil flow in between the flocs is probably much more complex than simple shear.

\subsubsection{Rotational geometries.~~\label{s.plate_images}}

\begin{figure}[htb]
\begin{center}
\scalebox{1.0}{\includegraphics{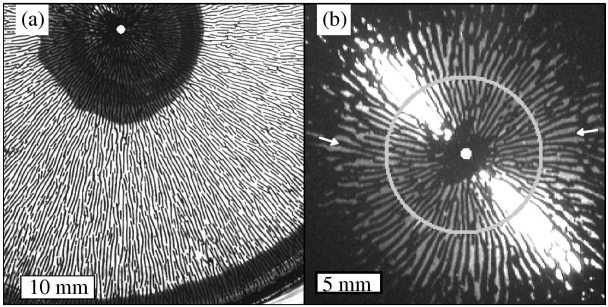}}
\end{center}
\caption{\small{Shear-induced structures in rotational geometries for $C=2$~\% w/w and $\gp=0.5$~s$^{-1}$. (a) Plate-plate geometry with $h=250~\mu$m. (b) Cone-and-plate geometry. The gray circle shows the limit between the conical and the plane region of the upper cone of truncation 209~$\mu$m. The white arrows point to conical flocs. The white dots in (a) and (b) indicate the rotation axis.}}
\label{f.plate_images}
\end{figure}

Figure~\ref{f.plate_images} shows pictures of the gel after shear in the rotational plate-plate and cone-and-plate geometries. The qualitative features of the pattern of vorticity-aligned flocs are in full agreement with previous observations \cite{Osuji:2008a,Negi:2009a}. In the plate-plate setup [see Fig.~\ref{f.plate_images}(a)], the wavelength seems to be constant in space but a large number of defects are seen due to the curved geometry.

In the cone-and-plate geometry [see Fig.~\ref{f.plate_images}(b)], the two large white spots correspond to reflections of the incident light on the metallic cone. Due to the rather large truncation of our cone, a circle was drawn to clearly differentiate between the inner plane region (for $r<6$~mm) and the conical region (for $r>6$~mm) of the tool. In this geometry, the shear-induced structures are wider at the cone periphery and thinner at the center. In some parts of the image, conical flocs, whose apparent width increases with the distance from the center, can be clearly identified [see arrows in Fig.~\ref{f.plate_images}(b)]. Although this result may seem natural, this shows unambiguously that flocs are not necessarily cylindrical and that their morphology is governed by the geometry of the shear cell.

\subsection{Quantitative results from image analysis \label{s.analysis}}

\subsubsection{Advection speed of the vorticty-aligned flocs.~~\label{s.speed}}

\begin{figure}[htb]
\begin{center}
\scalebox{1.0}{\includegraphics{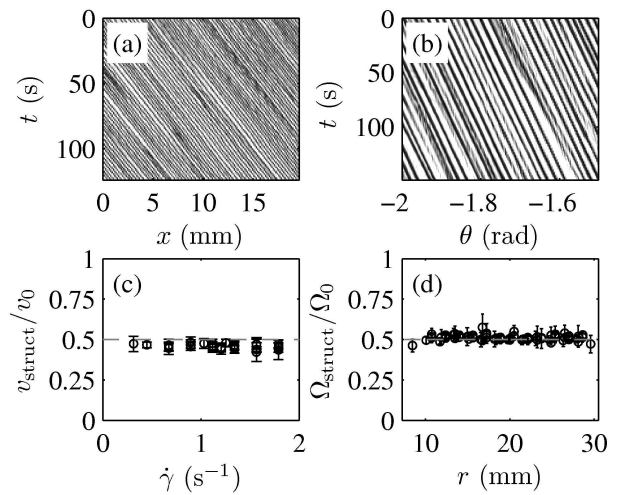}}
\end{center}
\caption{\small{Spatio-temporal diagrams of (a) a horizontal section of Fig.~\ref{f.simple_images} and (b) a circle of radius $r=20$~mm in Fig.~\ref{f.plate_images}(a). (c) Advection speed $v_{\rm struct}$ of the shear-induced structures normalized by the velocity $v_0$ of the bottom plate as a function of the shear rate $\gp$ in the translational setup for various concentrations ($C=1$--2.5~\% w/w) and $h=173~\mu$m. (d) Rotation speed $\Omega_{\rm struct}$ of the shear-induced structures normalized by the rotation speed $\Omega_0$ of the upper plate as a function of the radial position $r$ in the plate-plate rotational setup for various shear rates ($\gp=0.5$--5~s$^{-1}$), various gap widths ($h=100$--400~$\mu$m), and $C=2$~\% w/w. The gray dashed lines correspond to 0.5.}}
\label{f.speed}
\end{figure}

Figures~\ref{f.speed}(a) and (b) present spatio-temporal diagrams extracted from pictures taken in the translational shear cell and in the plate-plate rotational geometry during the steady-state regime. Such diagrams are obtained by plotting a horizontal section of the image in the translational case and a circle centered on the rotation axis in the rotational case as a function of time. The time origin $t=0$ is arbitrarily chosen in the steady-state regime. Both diagrams consist of parallel black and white stripes that reflect the motion of single structures at a constant speed. By measuring the slope of these stripes  in either case, one can easily estimate the floc advection speed $v_{\rm struct}$ [Fig.~\ref{f.speed}(c)] or their rotation speed $\Omega_{\rm struct}$ [Fig.~\ref{f.speed}(d)]. In both cases, the speed of the structures is always half that of the upper plate whatever the shear rate, the position in the sample, the gap width, and the gel concentration. This result indicates a solid-body rotation of cylindrical flocs between the two plates. As already noted above, this picture is directly confirmed by movies\dag~recorded at a large magnification in the translational setup.

\subsubsection{Wavelength and floc width extraction.~~}

\begin{figure}[htb]
\begin{center}
\scalebox{1.0}{\includegraphics{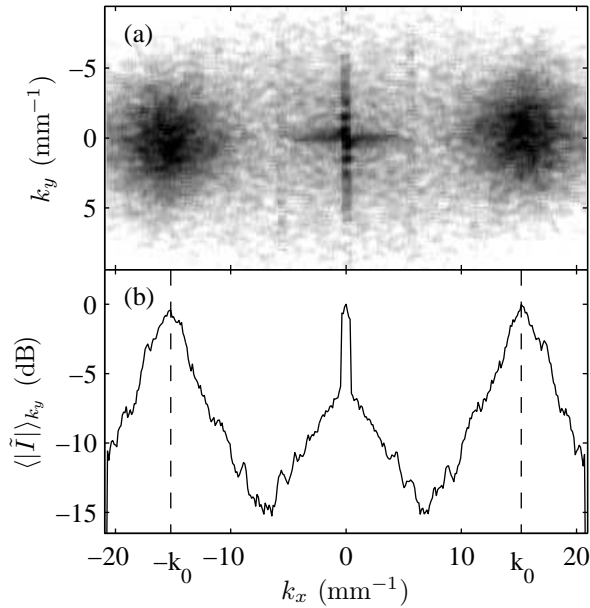}}
\end{center}
\caption{\small{(a) Amplitude $|\tilde{I}(k_x,k_y)|$ of the 2D fast Fourier transform taken over the dotted frame shown in Fig.~\ref{f.simple_images} and filtered as explained in the text. (b) Average of the 2D FFT shown in (a) over the $k_y$-axis for $-0.6\lesssim k_y\lesssim 0.6$~mm$^{-1}$. The wavelength is readily extracted as $\lambda=2\pi/k_0$, where $k_0$ is the wavenumber corresponding to the maximum away from the origin (see dashed lines). The FFT amplitude is coded in decibels with the reference taken at the maximum i.e. we plot $20\log_{10}(|\tilde{I}(k_x,k_y)|/\max(|\tilde{I}|)$. In (a), black corresponds to 0 dB and white indicates levels smaller than -20~dB.}}
\label{f.image_analysis}
\end{figure}

The wavelength $\lambda$ of the shear-induced pattern is extracted from the images shown above by Fourier transform. In the case of the translational shear cell, we first compute the 2D fast Fourier transform (FFT) $\tilde{I}(k_x,k_y)$ of a given grayscale image $I(x,y)$ over a small area of typical size 3~cm$^2$ (see the dotted frame in Fig.~\ref{f.simple_images}). This raw FFT is then filtered using a 2D convolution over 3 points (7 resp.) in the $k_y$ ($k_x$ resp.) direction in order to remove the noise [see Fig.~\ref{f.image_analysis}(a)]. Finally, we average the 2D FFT for $-0.6 \lesssim k_y\lesssim 0.6$~mm$^{-1}$ and we look for the maximum of $\left < |\tilde{I}(k_x,k_y)| \right > _{k_y}$ away from the origin and for the corresponding $k_x=k_0$ to extract the wavelength $\lambda=2\pi /k_0$. Using the larger magnification (see inset of Fig.~\ref{f.simple_images}), we also measure directly the apparent floc width $d$ under simple shear. By selecting various parts of the sample, we checked that both $\lambda$ and $d$ do not vary in space (data not shown).

\begin{figure}[htb]
\begin{center}
\scalebox{1.0}{\includegraphics{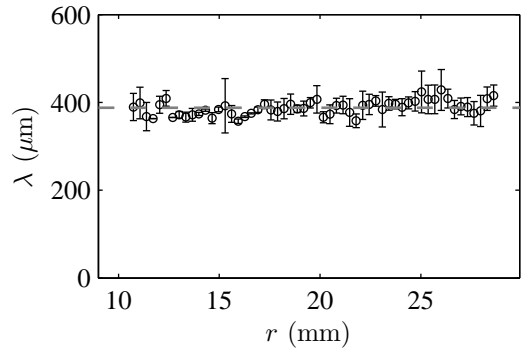}}
\end{center}
\caption{\small{Wavelength $\lambda$ of the shear-induced pattern as a function of the radial distance $r$ in the plate-plate rotational geometry. $C=2$~\% w/w, $h=150~\mu$m, and $\gp=0.5$~s$^{-1}$. The dashed line shows the mean value of $\lambda$ over all values of $r$.}}
\label{f.plate_r}
\end{figure}

In the case of the rotational setup, from images similar to Fig.~\ref{f.plate_images}, we extract the intensity $I(r,\theta)$ as a function of the angle at a fixed distance $r$ from the rotation axis. Then we calculate the FFT of $I(r,\theta)$ over $\theta$, whose maximum yields the angular period that we multiply by the distance to the rotation axis to recover the spatial period $\lambda(r)$ of the pattern. As for the translational setup, the floc width $d$ is extracted by directly measuring the apparent width of vorticity-aligned rolls from calibrated images obtained under large magnification. 

In all cases $\lambda$ and $d$ are computed for 20 to 100 images recorded in the steady state regime. The values shown in the following figures are the average of these measurements and the error bars show the standard deviation of these measurements. As an example, the wavelength measured in the plate-plate rotational device for various values of $r$ is presented in Fig.~\ref{f.plate_r}. These data show that $\lambda$ is spatially homogeneous so that what follows an additional spatial average over $r=10$--25~mm will be used when analyzing the experiments performed in the plate-plate geometry. 

\subsubsection{Influence of the shear rate.~~\label{s.gamma_dot}}

\begin{figure}[htb]
\begin{center}
\scalebox{1}{\includegraphics{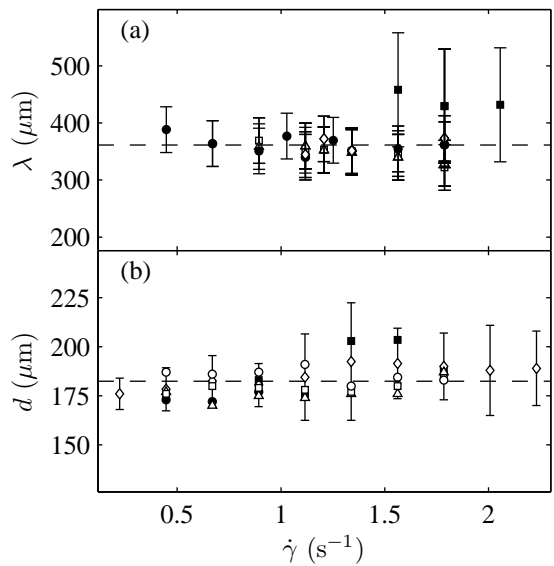}}
\end{center}
\caption{\small{Influence of the shear rate $\gp$. (a) Wavelength $\lambda$ of the shear-induced pattern and (b) apparent floc width $d$ vs $\gp$ in the translational shear cell ($h=173~\mu$m). The different symbols correspond to various concentrations: $C=0.5$~\% ($\circ$), 1~\% ($\square$), 1.5~\% ($\triangle$), 2~\% ($\bullet$), 2.5~\% ($\diamond$), and 3~\% w/w ($\blacksquare$). The dashed lines show the mean values of $\lambda$ and $d$ over all values of $\gp$ and $C$.}}
\label{f.gamma_dot}
\end{figure}

Figure~\ref{f.gamma_dot} presents the wavelength $\lambda$ and the floc width $d$ as a function of the applied shear rate $\gp$ for various concentrations in the translational setup. Up to experimental uncertainty, $\lambda$ and $d$ are independent of $\gp$ over one decade in shear rate. This means that some shear strain is necessary to induce the structures but that the rate at which this strain is applied is not relevant to the steady-state regime.

In the plate-plate rotational geometry, we recall that $\lambda$ does not depend on the radial distance $r$ (see Fig.~\ref{f.plate_r}). This also implies that $\lambda$ is independent of the local shear rate $\gp(r)$. By systematically varying the plate rotation speed we found that $\lambda$ does not depend on the global shear rate $\gp$, i.e. the shear rate at the periphery of the plate, either (data not shown).

Moreover, there clearly exists a critical shear rate $\gp_c$ above which structuration does not occur. We shall discuss the measurements of $\gp_c$ and the resulting ``phase diagram'' in more details in Sect.~\ref{s.diagphi} below.

\subsubsection{Influence of the concentration.~~\label{s.concentration}}

\begin{figure}[htb]
\begin{center}
\scalebox{1.0}{\includegraphics{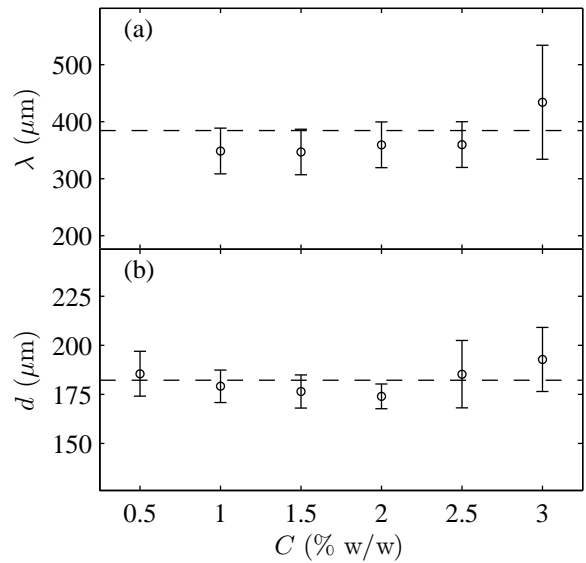}}
\end{center}
\caption{\small{Influence of the gel concentration $C$. (a) Wavelength $\lambda$ of the shear-induced pattern and (b) apparent floc width $d$ vs $C$ in the translational shear cell ($h=173~\mu$m). Each data point corresponds to an average over the various shear rates shown in Fig.~\ref{f.gamma_dot}. The dashed lines show the mean values of $\lambda$ and $d$ over all values of $C$.}}
\label{f.concentr}
\end{figure}

As seen in Fig.~\ref{f.concentr} for the translational setup, both the wavelength and the floc width do not show any clear dependence on the concentration $C$. The same result was obtained in the plate-plate rotational geometry (data not shown). Interestingly, this suggests that the mechanism underlying the structure formation and selecting the pattern is insensitive to the gel concentration. It also clearly implies that the floc density has to increase with increasing concentration, which we shall discuss further in Sect.~\ref{s.discuss}. Above a concentration of 3~\% w/w, CB gels were not observed to present any clear shear-induced structuration in our range of gaps and shear rates.

\subsubsection{Influence of the gap width.~~\label{s.gap}}

\begin{figure}[htb]
\begin{center}
\scalebox{1}{\includegraphics{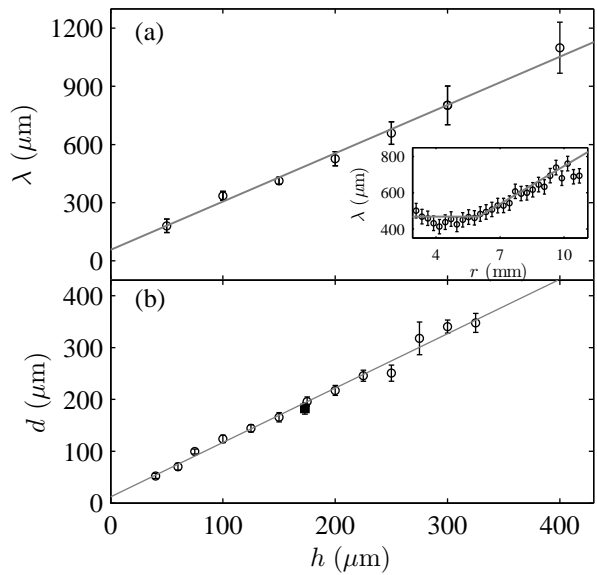}}
\end{center}
\caption{\small{Influence of the gap width $h$. (a) Wavelength $\lambda$ of the shear-induced pattern vs $h$ for $C=2$~\% w/w in the plate-plate rotational geometry. The solid line is $\lambda=2.5\, h+60~\mu$m. Inset: $\lambda$ vs the radial distance $r$ for $C=2$~\% w/w in the cone-and-plate geometry. For $r>6$~mm, the solid line is $2.0\, h(r)+50~\mu$m, where $h(r)=r\theta$ is the local gap and $\theta=2^\circ$ is the cone angle. For $r<6$~mm, the gap width is constant and equal to $h=209~\mu$m due to the cone truncation. (b) Apparent floc width $d$ vs $h$ in the plate-plate rotational geometry for $C=2$~\% w/w ($\circ$). The black square ($\blacksquare$) shows the floc width measured in the translational shear cell ($h=173~\mu$m). The solid line is $d=1.05\, h+12~\mu$m.}}
\label{f.gap}
\end{figure}

As shown in Fig.~\ref{f.gap}(a) for a 2~\% w/w gel in the plate-plate rotational geometry, $\lambda$ strongly depends on the gap width $h$. A linear fit yields $\lambda=2.5\, h + 60~\mu$m. This linear scaling is confirmed in the inset of Fig.~\ref{f.gap} through experiments performed in the cone-and-plate geometry. In this geometry both the gap and the pattern wavelength are not uniform. However, when considering the local gap $h(r)$ and the local wavelength $\lambda(r)$, one recovers a linear relationship $\lambda(r)=2.0\, h(r)+50~\mu$m, which is consistent with the previous ones. We conclude that the average linear law $\lambda\simeq 2.25\, h+55~\mu$m provides a good description of the data whatever the geometry, the gel concentration, and the applied shear rate. Still, due to experimental uncertainty, it remains unclear from these data whether the pattern wavelength tends to zero or not under extreme confinement $h\rightarrow 0$.

Finally, Fig.~\ref{f.gap}(b) shows that the floc width $d$ also increases linearly with the gap width: $d=1.05\, h+12~\mu$m. Here, the y-intercept of 12~$\mu$m is believed to be insignificant in view of the experimental uncertainty. Therefore, our results indicate that the apparent floc width $d$ coincides with the gap width $h$, as can also be checked in the translational setup in Figs~\ref{f.gamma_dot}, \ref{f.concentr}, and \ref{f.gap}(b).

\subsubsection{Phase diagram for shear-induced structuration.~~\label{s.diagphi}}

\begin{figure}[htb]
\begin{center}
\scalebox{1.0}{\includegraphics{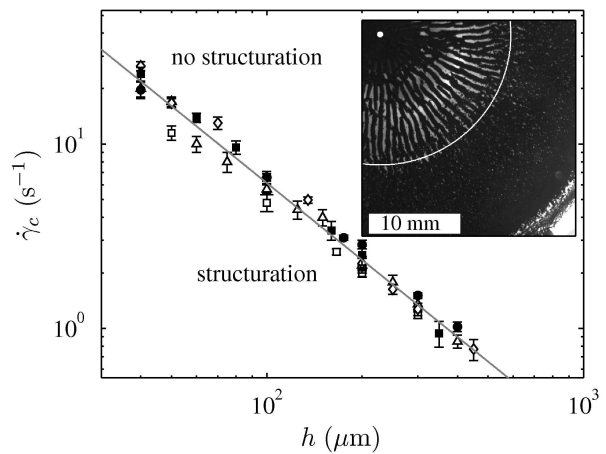}}
\end{center}
\caption{\small{Critical shear rate $\gp_c$, above which no structuration is observed, as a function of the gap width $h$ and for various concentrations: 1~\% ($\square$), 1.5~\% ($\triangle$), 2~\% ($\bullet$), 2.5~\% ($\diamond$), and 3~\% w/w ($\blacksquare$). The solid line is the best power-law fit of the full data set: $\gp_c=A\,h^{-\alpha}$ with $A=3650$ and $\alpha=1.39$. Inset: determination of $\gp_c$ in the rotational plate-plate setup for $C=2$~\% w/w, $h=300~\mu$m, and $\gp=2.5$~s$^{-1}$ (see text). The white dot indicates the rotation axis and the white circle shows the critical radius $r_c$ beyond which no structuration is observed.}}
\label{f.diagphi}
\end{figure}

As already noted, shear-induced structuration does not occur when the applied shear rate exceeds a characteristic shear rate $\gp_c$. A simple way to measure $\gp_c$ is to take advantage of the non-uniformity of the shear rate in the rotational plate-plate setup, where the local shear rate $\gp(r)$ spans the whole range from 0 to the shear rate $\gp$ at the periphery of the plate. Indeed, if one imposes a large enough $\gp$, one observes that vorticity-aligned rolls develop only in the inner part of the sample while the outer part always remains homogeneous as seen in the inset of Fig.~\ref{f.diagphi}. This allows for a precise measurement of the critical radius $r_c$ that separates structuration from no structuration. $\gp_c$ is then defined as $\gp_c=\gp(r_c)=\gp r_c/R$. We checked that different estimates of $r_c$ extracted from images at various $\gp$ at the periphery yield the same values of $\gp_c$ to within 10~\%.

Figure~\ref{f.diagphi} shows $\gp_c$ plotted in logarithmic scales as a function of the gap width $h$ for different concentrations. Here again, the data does not depend significantly on the concentration. The resulting curve, which separates the homogeneous regime from the structuration domain in the $(\gp,h)$ ``phase diagram,'' is well fitted by a power-law behaviour $\gp_c=A\,h^{-\alpha}$ with $\alpha=1.4\pm 0.1$ 

\section{Discussion and conclusion\label{s.discuss}}


In summary, the main results of the present work are as follows. The steady-state features of shear-induced vorticity-aligned rolls are independent of the imposed shear rate $\gp$ (provided $\gp$ is smaller than a well-defined critical shear rate $\gp_c$) and of the concentration $C$ of the CB particles. Whatever the shearing geometry, the only parameter that controls the shear-induced pattern is the gap width $h$ and we found linear dependences $d\simeq h$ and $\lambda\simeq 2.25\,h$ for the apparent floc width $d$ and for the pattern wavelength $\lambda$ respectively. Finally, the critical shear rate $\gp_c$ is also independent of $C$ and scales as $h^{-1.4}$.

This last scaling may be explained in terms of a simple force balance between the viscous drag force acting on a floc and the attractive forces that keep individual particles together. Such a force balance is classically used to estimate the typical cluster size as a function of the shear rate in flocculated suspensions under shear \cite{Negi:2009a}. Here, we consider flocs generated from the inital gel network before structuration and elongation along the vorticity direction occurs. For confined CB gels, the typical size of such flocs is given by the gap width $h$ so that the viscous force can be estimated as $F_{\rm visc}\sim \eta_s\gp h^2$, where $\eta_s$ is the oil viscosity. This viscous force will be able to tear the floc apart, say in two pieces of size $h/2$, if it exceeds the attractive force $F_{\rm attr}\sim N_s U/\delta$, where $\delta$ is the range of the attractive potential, $U$ its depth, and $N_s$ is the number of CB particles in a cross-section of the floc. Introducing the floc fractal dimension $d_f$, one has $N_s\sim h^{d_f-1}$ so that $F_{\rm attr}\sim h^{d_f-1} U/\delta$. Therefore, structuration into flocs of size $h$ is predicted to be prevented by viscous drag for $F_{\rm visc}>F_{\rm attr}$, which corresponds to $\gp >\gp_c$ with
\begin{equation}
\gp_c\sim \frac{U}{\eta_s\delta}\,h^{d_f-3}\,.
\label{e.gpc}
\end{equation}

\begin{figure}[!ht]
\begin{center}
\scalebox{2.2}{\includegraphics{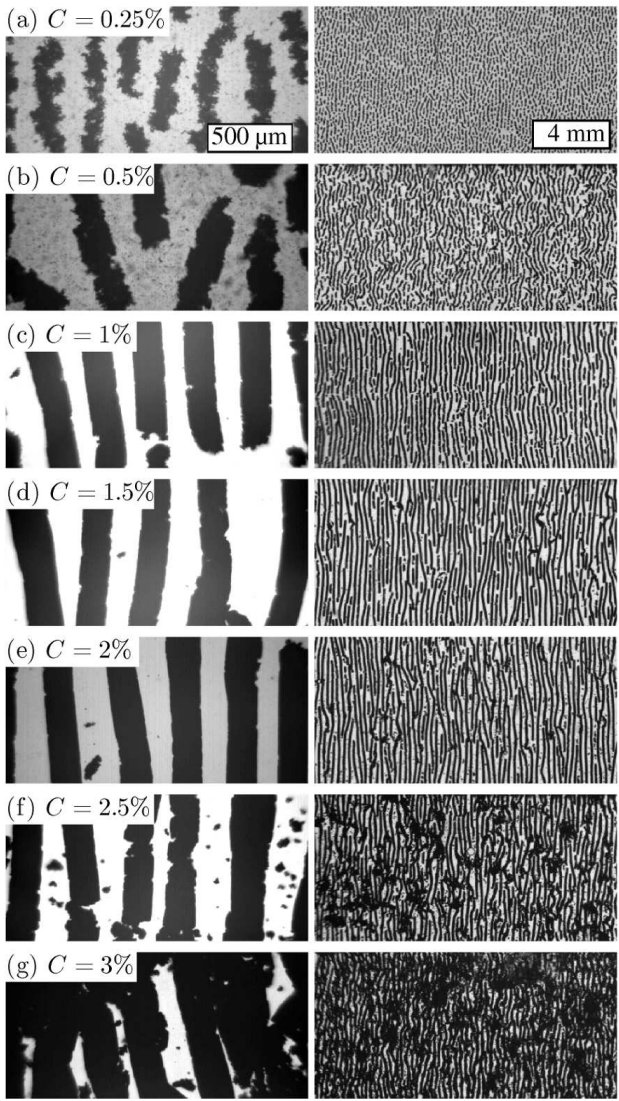}}
\end{center}
\caption{\small{Shear-induced patterns under simple plane shear with $h=173~\mu$m for different gel concentrations $C$. Left: large magnification. Right: low magnification.}}
\label{f.imphi}
\end{figure}

The critical shear rate is thus predicted to follow a power-law behaviour and, in the framework of Eq.~\eqref{e.gpc}, we deduce $d_f=3-\alpha=1.6\pm 0.1$ from experimental observations. Previous works devoted to the equilibrium structure of different CB particles (including the ones used in the present study) suspended into various oils have reported fractal dimensions $d_f=1.8\pm 0.1$, hinting at a Diffusion Limited Cluster Aggregation (DLCA) mechanism \cite{Ehrburger-Dolle:1990,Bezot:1998}. However, such an agreement between the fractal dimension of the gel network at rest and that deduced from the experimental measurements of $\gp_c$ should be mitigated since shear-induced flocs are not expected to keep the same fractal dimension as the intial gel state. Indeed, both experimental and theoretical studies of weakly aggregated dispersions have shown an increase of the fractal dimension from 1.8 to about 2.3 due to shear \cite{Sonntag:1986,Potanin:1995,Thill:1998}. Therefore, the approximations leading to Eq.~\eqref{e.gpc} may be too crude and a more subtle theoretical approach, such as that developed by Potanin,\cite{Potanin:1991} would probably be better suited. In any case, it would be interesting to also test the influence of $U$ and $\eta_s$, e.g., by adding a dispersant to lower the interaction potential.

While the fact that the floc size is fixed by the gap width seems rather natural for confined log-rolling aggregates, a surprising result of the present study is that all the steady-state features of the vorticity-aligned rolls, namely $\lambda$, $d$, and $\gp_c$, are independent of the gel concentration $C$.
If one assumes a perfect pattern of cylindrical rolls of diameter $h$ with a wavelength $\lambda$, it is easily shown that the particle volume fraction within a roll $\varphi_{\rm roll}$ is linked to the initial volume fraction of the gel at rest $\varphi$ by
\begin{equation}
\varphi_{\rm roll}=\frac{4\lambda}{\pi h}\,\varphi\simeq\frac{4\lambda}{\pi h}\,\frac{\rho_{\rm oil}}{\rho_{\rm CB}}\,C\,,
\label{e.phi}
\end{equation}
where $\rho_{\rm oil}$ and $\rho_{\rm CB}$ denote the densities of the suspending oil and of the CB particles respectively and the last approximation results from $\varphi=\rho_{\rm oil}C/[\rho_{\rm oil}C+\rho_{\rm CB}(1-C)]$ with $C\ll 1$. Since the wavelength $\lambda$ does not depend on $C$, Eq.~\eqref{e.phi} shows that the volume fraction inside vorticity-aligned rolls is simply proportional to $C$. Enlarged images of the shear-induced structures at various concentrations confirm qualitatively that the flocs are very tenuous at the lowest concentration [$C=0.25$~\% w/w, see Fig.~\ref{f.imphi}(a)] and get more and more compact as $C$ is increased [see Fig.~\ref{f.imphi}(b-c)]. 


Still, when taking a wider look at the shear-induced pattern (see right column of Fig.~\ref{f.imphi}), it clearly appears that the picture of an ideal striped pattern used in Eq.~\eqref{e.phi} is not very realistic outside the range $C=1$--2~\% w/w. Indeed, Fig.~\ref{f.imphi}(a-b) shows that lots of ``holes'' are present in the shear-induced pattern obtained at the lowest concentrations (0.25--0.5~\% w/w) while at the largest concentrations (2.5--3~\% w/w), very large agglomerates superimpose to the roll pattern [see Fig.~\ref{f.imphi}(f-g)]. This suggests that there exists a maximum volume fraction for the vorticity-aligned structures above which the rolls can no longer compress. However, in this case, it is unclear why the system does not lower the pattern wavelength to accommodate more rolls per unit area. Rather than a decrease in $\lambda$, we observe a coexistence between very large clusters and the same roll pattern.

To conclude, the present experimental data set paves the way for future modelling of shear-induced vorticity alignment in attractive systems. In particular, we believe that theoretical works should concentrate on identifying the mechanism for structure formation and on the prediction of the pattern wavelength. Such a prediction may result not only from  considerations on viscoelastic effects within the gel as already suggested in previous studies~\cite{Lin-Gibson:2004,Montesi:2004} but also from a detailed analysis of the oil flow around the flocs. Moreover, dynamical information on the pattern formation based on both image analysis and transient rheology should provide even more experimental input to such theoretical attempts.


\begin{acknowledgments}

We thank D. Tamarii for technical help with the rheometer and L. Cipelletti, T. Divoux, T. Gibaud, and V. Trappe for fruitful discussions. V. Trappe is also thanked for providing us with the carbon black powder.

\end{acknowledgments}




\end{document}